\documentclass[11pt,a4paper]{article} 
\pdfoutput=1
\usepackage{jheppub}

\usepackage{graphicx}
\usepackage{epstopdf}
\usepackage{amsmath, amsthm, amssymb}
\usepackage{mathrsfs}
\usepackage{amssymb}
\usepackage{verbatim}
\usepackage{color}
\usepackage{multirow}
\usepackage{ulem}

\newcommand{\beq}{\begin{equation}}
\newcommand{\eeq}{\end{equation}}
\newcommand{\bea}{\begin{eqnarray}}
\newcommand{\eea}{\end{eqnarray}}

\newcommand{\gsim}{\lower.7ex\hbox{$\;\stackrel{\textstyle>}{\sim}\;$}}
\newcommand{\lsim}{\lower.7ex\hbox{$\;\stackrel{\textstyle<}{\sim}\;$}}

\newcommand{\mrm}{\mathrm}
\newcommand{\df}{\dfrac}

\def\stacksymbols #1#2#3#4{\def\theguybelow{#2}
    \def\vp{\lower#3pt}
    \def\sp{\baselineskip0pt\lineskip#4pt}
    \mathrel{\mathpalette\intermediary#1}}

\def\intermediary#1#2{\vp\vbox{\sp
     \everycr={}\tabskip0pt
     \halign{$\mathsurround0pt#1\hfil##\hfil$\crcr#2\crcr
              \theguybelow\crcr}}}

\def\be{\begin{equation}}
\def\ee{\end{equation}}
\def\bea{\begin{eqnarray}}
\def\eea{\end{eqnarray}}

\def\sp{\;\;\;,\;\;\;}

\def\mrm{\mathrm}

\def\lsim{\raise0.3ex\hbox{$\;<$\kern-0.75em\raise-1.1ex\hbox{$\sim\;$}}}
\def\gsim{\raise0.3ex\hbox{$\;>$\kern-0.75em\raise-1.1ex\hbox{$\sim\;$}}}

\def\inbar{\,\vrule height1.5ex width.4pt depth0pt}

\def\IC{\relax\hbox{$\inbar\kern-.3em{\rm C}$}}
\def\IQ{\relax\hbox{$\inbar\kern-.3em{\rm Q}$}}
\def\IR{\relax{\rm I\kern-.18em R}}
 \font\cmss=cmss10 \font\cmsss=cmss10 at 7pt
\def\IZ{\relax\ifmmode\mathchoice
 {\hbox{\cmss Z\kern-.4em Z}}{\hbox{\cmss Z\kern-.4em Z}}
 {\lower.9pt\hbox{\cmsss Z\kern-.4em Z}}
 {\lower1.2pt\hbox{\cmsss Z\kern-.4em Z}}\else{\cmss Z\kern-.4em Z}\fi}

\def\comment#1{}
\def\to{\rightarrow}

\def\u1x{U(1)_X}
\newcommand{\nc}{\newcommand}
\nc{\LL}{L}
\nc{\vv}{\tilde{v}}
\nc{\ccdot}{\!\cdot\!}
\nc{\gsm}{G_{SM}}
\nc{\vfive}{\mathbf{5}\oplus\mathbf{\overline{5}}}
\nc{\vten}{\mathbf{10}\oplus\mathbf{\overline{10}}}
\nc{\zhol}{Z^{\rm hol}}
\nc{\xfb}{\,{\rm fb}}

\begin{document}

\vspace*{1mm}

\title{Invisible $Z'$ and dark matter: LHC vs LUX constraints}

\author{Giorgio Arcadi$^{a}$}
\emailAdd{arcadi@theorie.physik.uni-goettingen.de}
\author{Yann Mambrini$^{b}$}
\emailAdd{yann.mambrini@th.u-psud.fr}
\author{Michel H.G. Tytgat$^{c}$}
\emailAdd{mtytgat@ulb.ac.be}
\author{Bryan Zald\'\i var$^{c}$}
\emailAdd{bryan.zaldivar@ulb.ac.be}

\vspace{0.1cm}
 \affiliation{
${}^a$ 
 Institute for Theoretical Physics, Georg-August University Gottingen, Friedrich-Hund-Platz 1,
 Gottingen, D-37077 Germany
}

\affiliation{
${}^b$ Laboratoire de Physique Th\'eorique 
Universit\'e Paris-Sud, F-91405 Orsay, France
 }
 
 \affiliation{
${}^c$ 
Service de Physique Th\'eorique, Universit\'e Libre de Bruxelles, Boulevard du Triomphe, CP225, 1050 Brussels, Belgium
}

\abstract{ 

We consider a simple, yet generic scenario in which a new heavy $Z'$ gauge boson couples both to SM fermions and to dark matter. In this framework we  confront the best LHC limits on an extra gauge boson $Z'$ to the constraints on  couplings to dark matter from direct detection experiments. 
 {In particular}  we show that {the LHC searches for resonant production of dileptons} and the recent {exclusion limits}  obtained by the LUX collaboration {give complementary constraints}. {Together, they} impose strong  bounds on the invisible branching ratio
 and exclude a large part of the parameter space for generic $Z'$ models.
Our study  encompasses many possible $Z'$ models, including SSM, $E_6$-inspired or B-L scenario. 

}

\maketitle
%}

%%%%%%%%%%%%%%%%%%%%%%%%%%%%%%%%%%%%%%%%%%%%%%%%%%%%%%%%%%%%%%%%%
%%%%%%%%%%%%%%%%%%%%%%%%%%%%%%%%%%%%%%%%%%%%%%%%%%%%%%%%%%%%%%%%%
%%%%%%%%%%%%%%%%%%%%%%%%%%%%%%%%%%%%%%%%%%%%%%%%%%%%%%%%%%%%%%%%%

%\maketitle

%% The arXiv's use of hypertex conflicts with revtex4's use of
%% \tableofcontents in single column format. To avoid this problem,
%% Include a file OOREADME.xxx with the word nohypertex in it when
%% you submit to the arXiv.
%\tableofcontents

%\section{Introduction}\label{sec:introduction}
\setcounter{equation}{0}

%%%%%%%%%

%%%%%%%%%%%%%%%%%%%%%%%%%%%%%%%%%%%%%%%%%%%%%%%%%%%%%%%%%%%%%%%%%%%%%%

\section{Introduction}
\label{sec:introduction}

The addition of extra
%dark
 $U(1)'$ gauge symmetries, associated to new  $Z^\prime$ massive neutral gauge bosons,
is among the well-motivated extensions of the Standard Model (SM) (for a review see e.g.  \cite{Langacker:2008yv, Han:2013mra}). 
Extra $U(1)'$ are not only predicted by some Grand Unified Theories (GUT), like $E_6$ or higher rank simple groups, 
but also appear systematically in string-inspired braneworld constructions \cite{string}. 
As the  SM particles do not have to be charged under the extra $U(1)'$, it is possible
 that a dark matter (DM) candidate lies in the extended gauge sector. In such a framework, a $Z^\prime$ may act as 
 messenger between  the visible sector (which contains the SM particles) and  a hidden sector (to which DM belongs), as for instance in the kinetic mixing portal scenario
\cite{Holdom,Rizzo:1998ut} (alternative scenarios are discussed, e.g, in \cite{Chiang:2012qa,Chiang:2013kqa}) . It has been shown that kinetic mixing is severely constrained by direct \cite{kindirect} or indirect detection \cite{kinindirect} and that only
some combination with scalar sector can relax the bounds \cite{kinscalar}.

The LHC experiments have set strong bounds on the mass of a $Z'$ coupled to SM particles, $M_{Z'} \gtrsim$ 2.5 TeV 
depending on the model \cite{Chatrchyan:2012oaa,Aad:2012hf}. However, these studies
consider only couplings to quark or leptons, and the LHC constraints  would be (somewhat) relaxed if the $Z'$ had non-negligible
couplings to dark matter. At the same time the XENON100 \cite{Aprile:2012nq} and, more recently, LUX \cite{Akerib:2013tjd} collaborations 
have set strong limits on spin-independent (SI) elastic cross section of dark matter with nucleons, whereas the
PICASSO \cite{Archambault:2012pm}  and COUPP \cite{Behnke:2012ys} collaborations are constraining spin-dependent (SD) interactions.
 
Clearly, if the $Z'$ also couples to dark matter, there is an interesting complementarity between LHC searches of new $Z'$ gauge bosons and direct searches for dark matter.  Indeed, for a given ${Z^\prime}$ mass, the direct detection experiments set upper limits on its couplings to dark matter; the same couplings trigger
 invisible decay channels for the $Z'$, which make the constraints from LHC experiments weaker. 
%while the LHC experiments constraints are weaker if the $Z'$ decays invisibly. 
In the present paper we investigate to what extent 
the constraints from LHC and direct detection are compatible. Some LHC study of invisible $Z'$ in some specific frameworks can be found in \cite{zplhc}
whereas some direct detection analysis can be found in \cite{zpdirect}.

The paper is organized as follows. After a summary of the model under consideration and its effective
 low energy limit, we establish in section II the relation between the invisible branching ratio of a $Z'$ and the direct detection
 cross sections (spin-dependent and spin-independent).  
We combine the exclusion limits set by the direct detection experiments with the LHC bounds on $Z'$ mass and couplings in section III. Some prospects are discussed in the last section.

\section{Invisible $Z'$ decay rates and direct detection cross sections}
\label{sec:DD}

%===========================  THE MODEL   ==============================================

\subsection{The model}

While the existence of extra $U(1)'$ gauge symmetries may be motivated in various ways \cite{Langacker:2008yv}, we will try to be as 
generic as possible in the present work. 
To do so, we parametrize as follows the couplings of a $Z'$ gauge bosons to the SM quarks and leptons on one hand and 
to a dark matter particle $\chi$ on the other hand, which we take to be fermionic for concreteness\footnote{Our results are similar for a scalar dark matter candidate and will be reported in a future work.}, 
\bea
\Delta {\cal L} \supset 
&&\;
g_D \bar \chi \gamma^\mu \left( V_D^\chi - A_D^\chi \gamma^5 \right) \chi ~ Z'_\mu
\nonumber
\\
&&+\; g_D \sum_f \bar f \gamma^\mu \left( V_D^f - A_D^f \gamma^5 \right) f ~ Z'_\mu.
\label{Eq:lmicro}
\eea
Here $V_D^f$ and $V_D^\chi$ ($A_D^f$ and $A_D^\chi$) are the vectorial (axial) couplings for the SM fermions $f$
and DM particle $\chi$ respectively ($V_D^\chi=0$ if $\chi$ is a Majorana particle) and we have introduced an overall gauge coupling
 $g_D$. Since the gauge coupling $g_D$, as well as $V^{f,\chi}_D$ and $A^{f,\chi}$ are a priori free parameters, our formalism can
  then describe all the possible $Z'$ realisations in terms of this five quantities. It is anyway possible to fix the combinations
   $g_D V^f_D$ and $g_D A^f_D$ by specifying the $Z'$ model (e.g. SSM, $E_6$, B-L,...) as it is shown in table I. 
   The partial widths of the $Z'$ can be straightforwardly 
   determined as function of these four fundamental parameters as: 
 \bea
  \label{Eq:z'width}
  \Gamma^{i}_{Z'}
  &=&
  \frac{g_D^2 c_i}
   {12 \pi} M_{Z'}\sqrt{1 - \frac{4 m^2_i}{M_{Z'}^2}} 
 \\
 &\times&
 \left[ (V_D^i)^2 \left(1+ 2 {m_i^2 \over M_{Z'}^2} \right) + (A_D^i)^2 \left(1- 4{ m_i^2 \over M_{Z'}^2}\right ) \right]
 \nonumber
 \eea 
where $i=f$ or $\chi$,  $c_f$ is the number of color of the SM fermion $f$  and $c_\chi=1$.
\begin{figure}[t]
    \begin{center}
    \includegraphics[width=4.in]{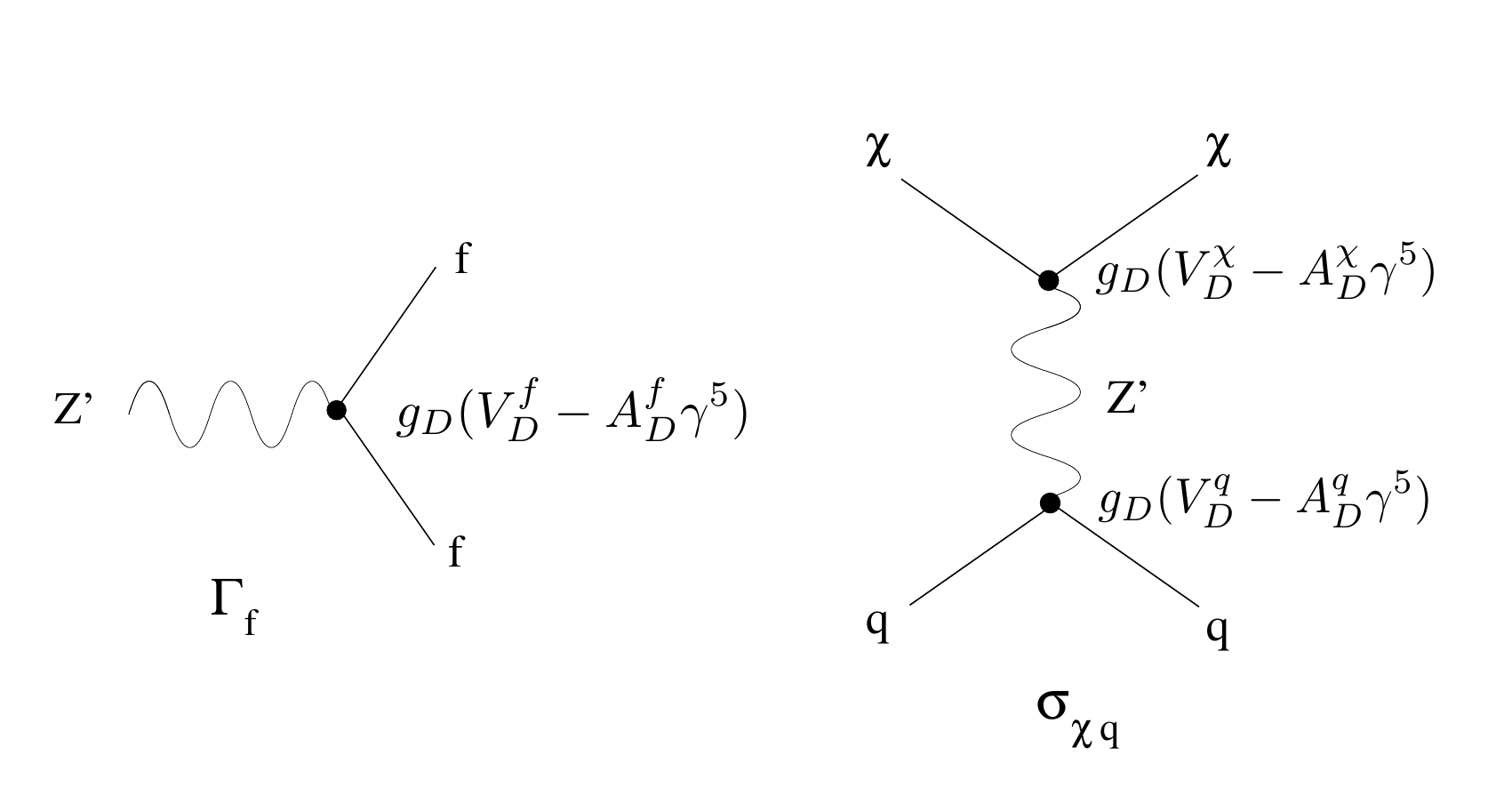}
              \caption{{\footnotesize
             Feynman  diagrams contributing to the width of the $Z'$ boson and the
             scattering cross section of the dark matter on a nucleon. 
}}
\label{Fig:feynman}
\end{center}
\end{figure}

From the microscopic Lagrangian (\ref{Eq:lmicro}), we can deduce the effective low energy interaction of the dark matter
to a nucleon. The XENON \cite{Aprile:2012nq} or 
LUX \cite{Akerib:2013tjd} experiments are supposed to observe dark matter collisions on nuclei,
which are translated as couplings to nucleons. These processes occur at velocities of the order of magnitude of that of the Sun 
in the galactic halo,
i.e. $\simeq 300$ km/s so that  the energy transferred in such  collisions is below the MeV scale 
for typical dark matter masses.
We can thus integrate out the $Z'$ from the Lagrangian (\ref{Eq:lmicro})  to get the effective Lagrangian
describing interactions between a $\chi$ and, say, a proton $p$, 
\bea
{\cal L}_{\mrm{eff}} 
&\supset&
\frac{g_D^2 }{M_{Z'}^2} \left[V_D^\chi (2 V_D^u + V_D^d ) \ \bar \chi \gamma^\mu \chi\, \bar p \gamma_\mu p\right.
\label{Eq:leff}
\\
&+&
\left. A_D^\chi (\Delta_u^p A_D^u +(\Delta_d^p	+ \Delta_s^p) A_D^d)
 \bar \chi \gamma^\mu \gamma^5 \chi\, \bar p \gamma_\mu \gamma^5 p\right],
\nonumber
\eea
with a similar expression for the neutron. 
Here $\Delta_i^p$ is the spin content of the quark $i$ in the proton, which may be extracted from lepton-proton scattering data. The
 values we use are taken from the latest determination of the light quark contributions \cite{Airapetian:2007mh}, i.e.
\be
\Delta_u^p=0.842,~~ \Delta_d^p = -0.427, ~~ \Delta_s^p = -0.085.
\ee

From  Eq.(\ref{Eq:leff}), we can deduce the Lagrangian describing the interaction 
of DM with a neutron by isospin symmetry\footnote{Another  description of this formalism can be found in \cite{dd}.}:
$2 V_D^u + V_D^d \rightarrow V_D^u + 2 V_D^d$ and $\Delta_u^n = \Delta_d^p$, $\Delta_d^n = \Delta_u^p$, 
$\Delta_s^n = \Delta_s^p$.
%\hspace{-1.cm}
\begin{table}
\caption{\footnotesize{Vectorial and axial couplings for some $Z'$ models.}}
\begin{center}
\begin{tabular}{|c|c|c|}
\hline
channels & $g_D~ V_{D}^f$ [SSM] & $g_D ~ A_D^f$ [SSM] \\
 \hline
$e^+~e^-$ & $\frac{g}{4 \cos \theta_W}(4 \sin^2 \theta_W -1)$ & $-\frac{g}{4 \cos \theta_W}$   \\
 \hline
 $\nu ~\nu$ & $\frac{g}{4 \cos \theta_W}$ &  $\frac{g}{4 \cos \theta_W}$ \\
\hline
 $u ~u $ & $\frac{g}{4 \cos \theta_W}(-\frac{8}{3} \sin^2 \theta_W +1) $ & $\frac{g}{4 \cos \theta_W}$   \\
 \hline
 $d ~d $ & $\frac{g}{4 \cos \theta_W}(\frac{4}{3}\sin^2 \theta_W -1 )$ & $-\frac{g}{4 \cos \theta_W}$   \\
\hline
\hline
 & $g_D~ V_{D}^f$ [B-L] & $g_D ~ A_D^f$ [B-L] \\
 \hline
$e^+~e^-$ & $-\sqrt{5/6}\tan \theta\, g$ & 0   \\
 \hline
 $\nu ~\nu$ & $-\sqrt{5/6}\tan \theta\, g$ &  0 \\
\hline
 $u ~u $ & $\sqrt{5/6}\tan \theta\,{g}/{3}$ & 0   \\
 \hline
 $d ~d $ & $\sqrt{5/6}\tan \theta\,{g}/{3}$ & 0  \\
\hline
\hline
 & $g_D~ V_{D}^f [E_6]$ & $g_D ~ A_D^f [E_6]$ \\
 \hline
$e^+~e^-$ & 0 & $\sqrt{5}\tan \theta\, g$   \\
 \hline
 $\nu ~\nu$ & 0 &  $\sqrt{5}\tan \theta\, g$ \\
\hline
 $u ~u $ & 0 & $\sqrt{5} \tan \theta\, g$  \\
 \hline
 $d ~d $ & 0 & $\sqrt{5} \tan\theta\, g$   \\
\hline
\end{tabular}
\end{center}
\label{tab:schemes}
\end{table}
As already discussed above, the combinations $g_D V_D^f$ and $g_D A_D^f$ are fixed in a given fundamental theory. For illustration, we report
 in Table I their values in some popular realizations of $Z'$ models, while a more exhaustive list can be found e.g. in \cite{Han:2013mra}.

For the sake of concreteness,  we will mainly in this paper refer to the so-called sequential model (noted SSM in the sequel) \cite{Langacker:2008yv}, for which the $Z'$ has the
 same couplings to SM fermions as the SM $Z$ boson. A possible realization of the SSM appears in constructions with extra dimensions at the weak scale, 
 but in the present work we merely take it as a benchmark model. Other realizations like (B-L) or $E_6$ $Z'$ do not change drastically our conclusion. 
Having set the $Z'$ model, the two parameters left free are the $\chi-Z'$ couplings, namely $V_D^\chi$ and $A_D^\chi$. We will show in the 
next section that $V_D^\chi$  is severely constrained
by  SI direct detection experiments, like   XENON100 \cite{Aprile:2012nq} or LUX \cite{Akerib:2013tjd}. On the contrary, $A_D^\chi$ can only be tested by 
SD direct detection experiments, like PICASSO \cite{Archambault:2012pm},  which are much less constraining. It is thus convenient to define a parameter 
\be
\alpha = A_D^\chi/V_D^\chi
\label{Eq:alpha}
\ee
 which determines the nature
of the coupling of the $Z'$ with dark matter,  $\alpha=0$ corresponding to a purely vectorial coupling, 
whereas large $\alpha$ correspond to a mostly axial coupling.  Although  $\alpha$  is of order unity for a generic model, a vector-like DM
 candidate would have $\alpha = 0$ while $\alpha = \infty$ for a Majorana DM candidate.
For concreteness we will  present our results for some  definite values, covering the range $0\leq\alpha\leq1000$.

%=====================================  DIRECT DETECTION  ====================================

\subsection{Direct detection}

From (\ref{Eq:leff}) we can derive the elastic scattering cross section
of  $\chi$ with a nucleon $N$, which takes place through t-channel exchange of a $Z'$, as illustrated in Fig.\ref{Fig:feynman}. In the case of SI scattering
 we may write the { effective} nucleon scattering cross section as
\bea
&&
\sigma_{\chi N}^{\mrm{SI}}= \frac{4 g_D^4 (V_D^\chi)^2 \mu_{\chi N}^2}{\pi M_{Z'}^4}
\bigg[ V_D^u \left(1 + \frac{Z}{A}\right) + V_D^d \left(2 - \frac{Z}{A}\right)  \biggr]^2
\label{Eq:sigmasi}
\eea
where $Z$ and $A$ are the charge number and  mass number of of the nucleus that contains the nucleon and 
$\mu_{\chi N}$ is the nucleon-dark matter reduced mass. Notice that direct detection experiments 
give their limits on $\chi-N$ elastic scattering cross section  assuming isopsin symmetry, $\sigma_{\chi p} = \sigma_{\chi n}=\sigma_{\chi N}$, where 
$p,n$ stands for the proton and neutron respectively.
This explains the presence in Eq.(\ref{Eq:sigmasi}) of a factor that depends on the atomic number $Z$ and mass number $A$ of the given
 nucleus\footnote{This implies that in general $\sigma_{\chi p} \neq \sigma_{\chi n}\neq \sigma_{\chi N}$, where the latter is defined
  in Eq.(\ref{Eq:sigmasi}).}. Similarly for Spin Dependent (SD) elastic scattering we obtain
\bea
\sigma_{\chi N}^{SD} =
&&
 \frac{12 g_D^4 \mu_{\chi N}^2 |A_D^\chi|^2}{\pi M_{Z'}^4(S_p^A + S_n^A)^2}
\biggl[ A_D^u(\Delta_u^p S_p^A 
+ \Delta_d^pS_n^A) 
\nonumber
\\
&&
+ A_D^d \left( (\Delta_d^p + \Delta_s^p) S_p^A + (\Delta_u^p + \Delta_s^p)S_n^A  \right) \biggr]^2
\label{Eq:sigmasd}
\eea 
$S_p^A$ and $S_n^A$ being the proton and neutron contribution to the nucleus spin.

Combining Eqs.(\ref{Eq:z'width}), (\ref{Eq:sigmasi}) and (\ref{Eq:sigmasd}) we can eliminate the parameters
$V_D^\chi$ and $A_D^\chi$, and express the invisible branching
ratio of the $Z'$ as function of physical observables, like
the spin dependent and spin independent scattering cross 
sections:\footnote{In the expression (\ref{Eq:brsigma}) we have assumed $\{m_f, m_\chi\} \ll M_{Z'}$ for sake of simplicity. The approximation
is mostly valid in the parameter space we are interested in,  but the dependence on
 the mass of the SM fermions and of the DM is obviously taken into account in our numerical analysis. }

\bea
Br_\chi &=& \frac{\Gamma_{Z'}^\chi}{\Gamma_{Z'}^\chi +\sum_f \Gamma_{Z'}^f } 
\label{Eq:brsigma}
\\
&=&
\biggl[ 1+  \left( \frac{2 g_D^2 \mu_{\chi N}}{M_{Z'}^2 \sqrt{\pi}} \right)^2
\frac{\sum_f c_f[ |V_D^f|^2  + |A_D^f|^2]}
{\sigma_{\chi N}^{\mrm{SI}}/\alpha_{Z,A}^{\mrm{SI}} + \sigma_{\chi N}^{\mrm{SD}}/ \alpha_{Z,A}^{\mrm{SD}} }   \biggr]^{-1}
\nonumber
\eea
with
\bea
&&
\alpha_{Z,A}^{\mrm{SI}}\equiv \biggl[ V_D^u (1+ \frac{Z}{A}) + V_D^d (2- \frac{Z}{A}) \biggr]^2
\nonumber
\\
&&
\alpha_{Z,A}^{\mrm{SD}} \equiv \frac{4}{(S_p^A + S_n^A)^2} 
\biggl[ A_D^u (\Delta_u^p S_p^A + \Delta_d^p S_n^A) 
\nonumber
\\
&&
+ A_D^d \left( [\Delta_d^p + \Delta_s^p] S_p^A +[\Delta_u^p + \Delta_s^p] S_n^A  \right) \biggr]^2.
\eea

It is convenient to re-express $Br_\chi$ in terms of
$\alpha$ (cf. (\ref{Eq:alpha})),
\bea
Br_\chi=\biggl[ 1+  \left( \frac{2 g_D^2 \mu_{\chi N}}{M_{Z'}^2} \right)^2  
\frac{\tilde c_F\alpha^{\rm SI}_{Z,A}}{\pi (1+ \alpha^2)\sigma_{\chi N}^{\mrm{SI}}}
\biggr]^{-1}
\label{Eq:bralpha}
\eea
with $\tilde c_F=\sum_f c_f (|V_D^f|^2 + |A_D^f|^2 )$. The expression (\ref{Eq:bralpha}), which establishes a relation between
the direct detection cross section and the invisible branching ratio of the $Z'$, is our main point. It implies that the exclusion limits 
on $\sigma_{\chi N}^{\mrm{SI}}$ set by XENON100 or LUX give an upper 
limit on the dark coupling to the $Z'$
and thus  an { upper limit} to its invisible width. The model dependence  is entirely encapsulated in $V_D^f$
and $A_D^f$, which correspond to the SM fermions couplings to the extra $U(1)$ gauge group.

%=======================================   LHC   ============================================

\subsection{LHC limits}

In this section we reexamine the current limits established at the LHC. 
They rely on searches of dileptons \cite{Aad:2012hf,Chatrchyan:2012oaa} or dijet \cite{ATLAS:2012pu,Chatrchyan:2013qha} 
resonances, and monojet or single photons with missing transverse energy \cite{Aad:2013oja,Chatrchyan:2012me}. 
In this work we will mainly rely on the limits actually set by the LHC collaborations. We are assuming a generic setup in which the $Z'$ couples both to leptons and quarks; as a consequence, among resonance searches, the ones of dileptons turn to be the most constraining~\footnote{The are also realizations in which $Z'$ does not couple to quarks, like e.g. \cite{Feng:2012jn}. Our analyis does not apply to such models.}. 
Regarding the searches of missing transverse energy, we notice that most of the analysis of the LHC collaborations are based on an effective operator approach, namely assuming that new heavy degrees of freedom (the $Z'$ our case) are integrated out. This approximation may not be appropriate in general (see for instance~\cite{Fox:2012ru}). In our analysis we have estimated possible modification of the limits when the effective approach is not valid while more quantitative results will require a dedicated study. We will notice anyway that in the example considered below, the monojet searches give subdominant constraints.

The existence of an invisible $Z'$  branching ratio  weakens the current LHC limits. Indeed,
at the partonic level the cross section for $\sigma(pp\to Z'\to\ell\ell)$ has the following
 Breit-Wigner profile near the resonance, 
\bea
\sigma(q\bar q\to Z'\to \ell\ell) &\approx& \df{g_D^4}{12\pi}(|V^q|^2+|A^q|^2) (|V^\ell|^2+|A^\ell|^2) \nonumber\\
&\times&\df{s}{(s-M_{Z'}^2)^2+\Gamma_{Z'}^2 M_{Z'}^2}~,
\eea
where $\Gamma_{Z'}$ is defined in Eq.~(\ref{Eq:z'width}). For a narrow resonance, $\Gamma_{Z'} \ll M_{Z'}$, the Breit-Wigner peak can be approximated by a Dirac delta, 
\be
\lim_{\epsilon\to 0 } \df{\epsilon}{\epsilon^2+a^2} = \pi\delta(a), 
\ee
so that 
\bea
\sigma(q\bar q\to Z'\to \ell\ell) 
&\approx & \df{g_D^4}{12\pi}(|V^q|^2+|A^q|^2) (|V^\ell|^2+|A^\ell|^2) \nonumber\\
&\times&\df{M_{Z'}}{\Gamma_{Z'}}\pi\delta(s-M_{Z'}^2).
\eea
As $\Gamma_{Z'} \propto g_D^2$ we see that the resonant cross-section scales as $g_D^2$. Now, the LHC  exclusion limits do not take into account the possibility of decay of a $Z'$ into some invisible sector, i.e. 
it is assumed that $Br_\chi = 0$. To implement an invisible $Z'$ width, we first write the visible width as $\Gamma_{Z'}^{SM} = \sum_f \Gamma_{Z'}^f$, which is the quantity used by the experiments in reporting their limits. Then 
\be
\label{eq:scaling}
\df{M_{Z'}}{\Gamma_{Z'}} =  \df{M_{Z'}}{\Gamma_{Z'}^{SM}} \df{\Gamma_{Z'}^{SM}}{\Gamma_{Z'}} = \df{M_{Z'}}{\Gamma_{Z'}^{SM}} (1-Br_\chi),
\ee
with $\Gamma_{Z'} = \Gamma_{Z'}^\chi + \Gamma_{Z'}^{SM}$. 
Thus, if all other things are kept constant, the resonant cross section into dileptons is diminished by a factor of $(1-Br_\chi)$.
We will furthermore assume that $g_D$ may vary freely for a generic $Z'$ model  (while $g_D=g$ reproducing the ATLAS/CMS limits for the SSM). 
Taking into account the dependence  of the cross section in $g_D$ and a non-zero $Br_\chi$ in a generic $Z'$ model, this implies a simple rescaling of
 the resonant cross section \footnote{Notice that the narrow width approximation is not always appropriated for some $Z'$ realizations, like in particular the SSM, 
 since it may occur that the decay width of the $Z'$ exceeds the detector resolution. In such a case the detector effect can be taken into account by convoluting 
 the Breit-Wigner expression with a Gaussian factor~\cite{Accomando:2010fz}, which depends on the width of the $Z'$ and the detector resolution. As a 
 consequence the relation~(\ref{eq:scaling}) should be modified accordingly. We have taken into account this factor in our numerical analysis.}  %in Fig.\ref{Fig:ATLAS}:
\be
\sigma_{Z'll}\to \left({g_D\over g}\right)^2 \times (1-Br_\chi) \times \sigma_{Z'll}.
\ee

\begin{figure}[t]
    \begin{center}
    \includegraphics[width=4.in]{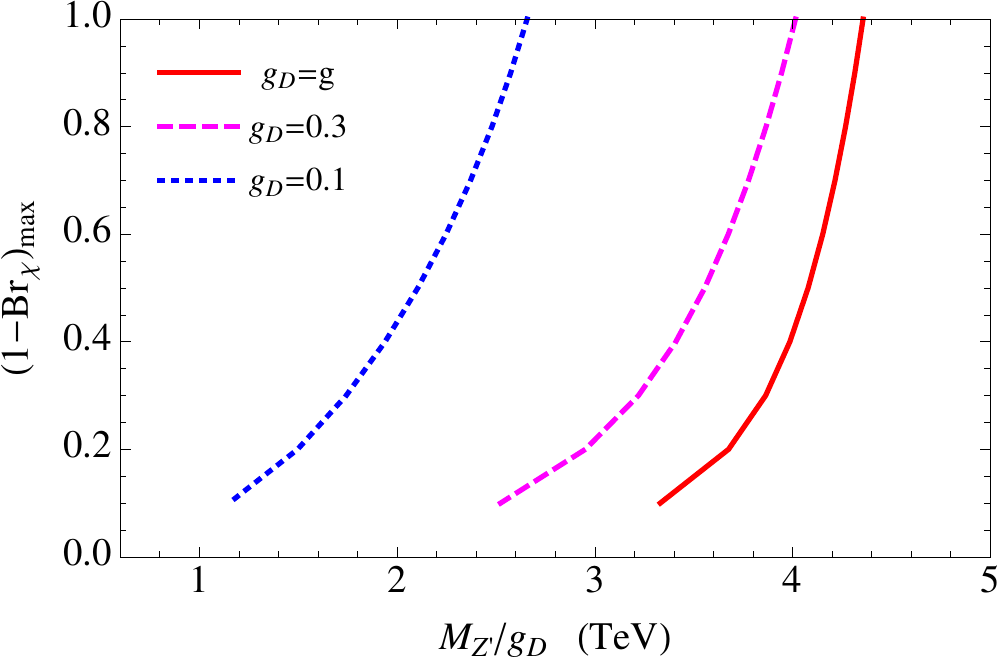}
              \caption{{\footnotesize
             Maximum visible branching ratio allowed respecting the limit obtained by ATLAS in the case of the SSM, for $g_D=g$ (red,solid), 0.3 (magenta, dashed) and 0.1 (blue, dotted) as the function
             of $M_{Z'}/g_D$.
}}
\label{Fig:br}
\end{center}
\end{figure}

In Fig.~(\ref{Fig:br}) we have applied this scaling relation to reinterpret the ATLAS exclusion limit in presence of an invisible decay width of the $Z'$. The new
 limits have been reported as function of the $Z'$ mass and for different values
of  $g_D$. The couplings of the $Z'$ with SM fermions have been assumed to be the ones of the SSM since this is, at the moment, the most 
constrained $Z'$ model. Thus  our limits can be regarded as being conservative.  
For instance, we can see that, for $g_D=0.3$, a $Z'$ with a mass of 800 GeV is
still allowed by the LHC data, provided its invisible branching ratio is large enough, $Br_\chi \gsim 40 \%$. 

\section{Results}

In order to confront the limits on an invisible $Z'$ from resonant dilepton production at the LHC to direct searches for dark matter, we  
concentrate in this work on spin independent scattering and in particular on the results from the LUX collaboration \cite{Akerib:2013tjd}.  The LUX detector,
 which  is operated at the Sanford Underground
Research Facility, is made of  a dual--phase xenon time--projection chamber.  With an exposure of $85.3$ live--days 
%$\times$ 118.3 kg fiducial volume,
 the LUX collaboration has set the strongest bounds on the SI elastic scattering cross section of dark matter, $\sigma^{\mrm{SI}}_{\chi N}$.   

\begin{figure}[t]
    \begin{center}
    \includegraphics[width=3.8in]{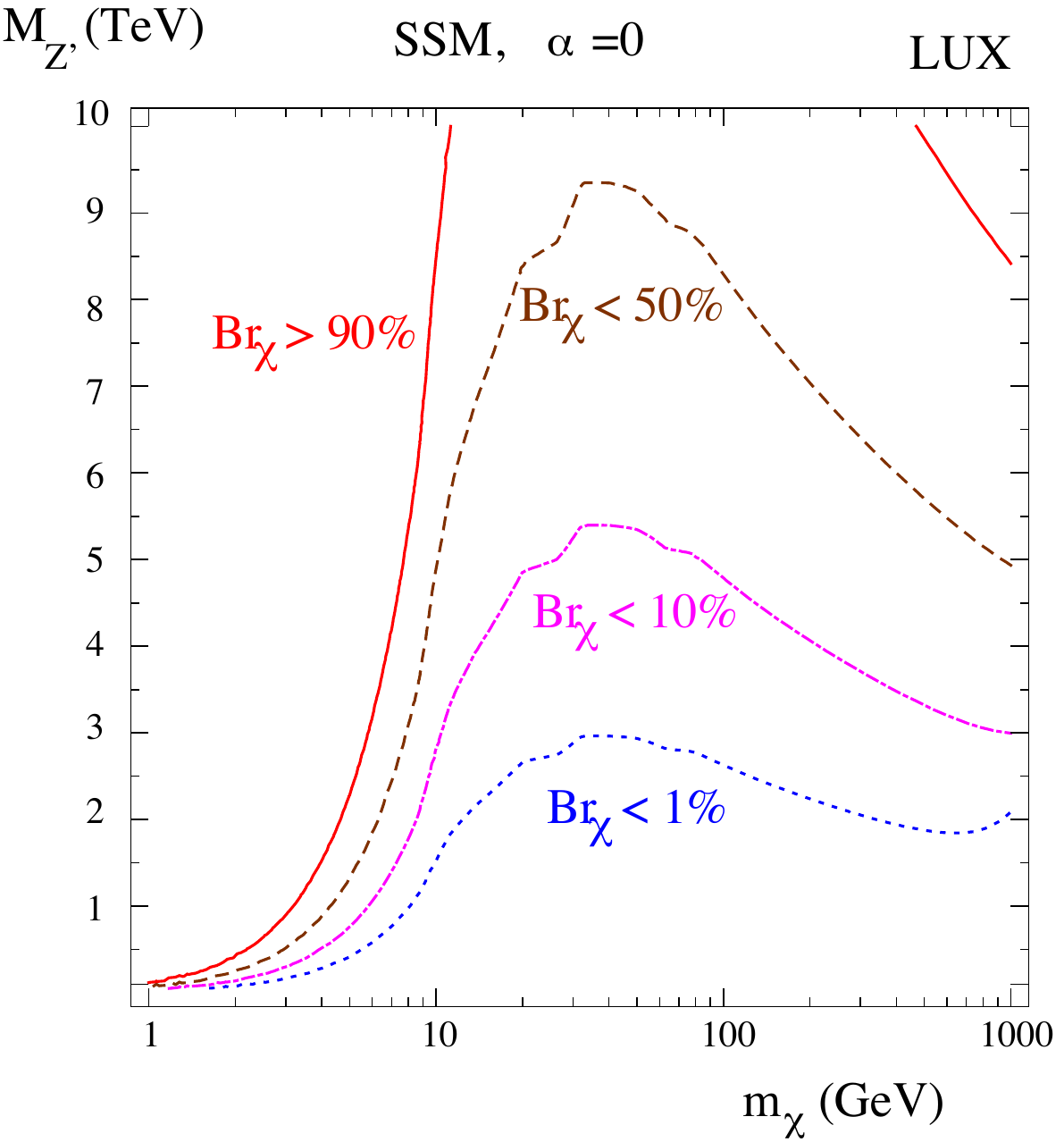}
              \caption{{\footnotesize
        Minimum values of $M_{Z'}$ according to iso-contours  of $Br_\chi$ as a function of DM mass, after applying LUX constraints
         in the case of  a pure vectorial coupling ($\alpha= A_D^\chi/V_D^\chi= 0$, see the text for details).     
         }}
\label{Fig:bralpha0}
\end{center}
\end{figure}

\begin{figure}[t]
    \begin{center}
    \includegraphics[width=3.8in]{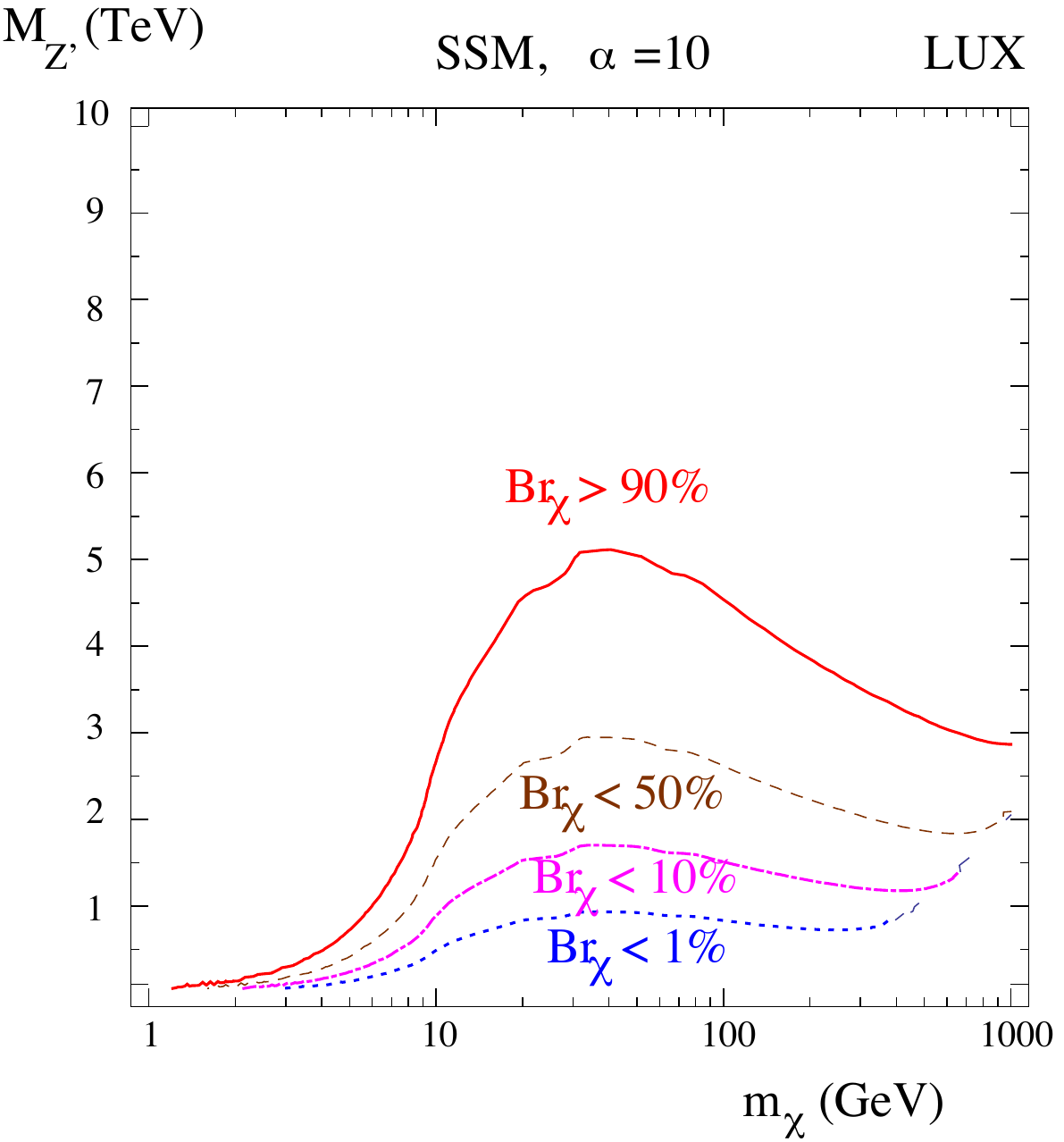}
              \caption{{\footnotesize
          Same as Fig. \ref{Fig:bralpha0} but with a large axial coupling, $\alpha=10$.     
          }}
\label{Fig:bralpha10}
\end{center}
\end{figure}

For a given dark matter candidate mass, we can translate the exclusion limit set by LUX experiment into
an upper bound on the invisible branching ratio of the $Z'$  using Eq.(\ref{Eq:bralpha}).
We present the result as limits in the plane ($m_\chi$, $M_{Z'}$) in figures (\ref{Fig:bralpha0}) and (\ref{Fig:bralpha10}), respectively  for 
for a purely vectorial coupling ($\alpha=0$) and for a mixture of vector and axial couplings ($\alpha=10$).

Fixing  $M_{Z'}$ and $m_\chi$, the figures give the allowed range of invisible branching ratios. For instance, for $M_{Z'} = 5$ TeV and $m_\chi = 100$ GeV, the LUX
 limits forbid an invisible branching ratio of the $Z'$ larger than 10 $\%$ for a purely vectorial coupling. 
Indeed,  a larger invisible branching ratio would mean a larger coupling of the $Z'$ with the 
 DM and thus a larger $\sigma^{\rm SI}_{\chi N}$. 
Clearly, for a heavier $Z'$, the invisible branching may be larger. 
Although the two plots share similar features, we notice that a non-zero axial couplings relaxes the limits on the constraints on the invisible 
branching ratio, since LUX is not very sensitive to such couplings which lead to spin dependent interactions. 
We further notice that very light dark matter candidate ($\lesssim 3$ GeV) can be consistent with an almost completely invisible light $Z'$ 
because the constraints set by LUX (and direct detection experiments) are much weaker in this case. On the contrary, for heavier dark matter candidates, 
and in particular for masses between roughly 50 and 100 GeV,
the limits on the invisible branching ratio are very severe, even for a quite heavy $Z'$. Indeed, this mass range corresponds to the region
where LUX is the most sensitive.
At the same time, if the invisible branching is small it should be easier to see the  $Z'$
 at the LHC. It is thus of interest to combine both analyses.

\begin{figure}[t]
    \begin{center}
    \includegraphics[width=4.3in]{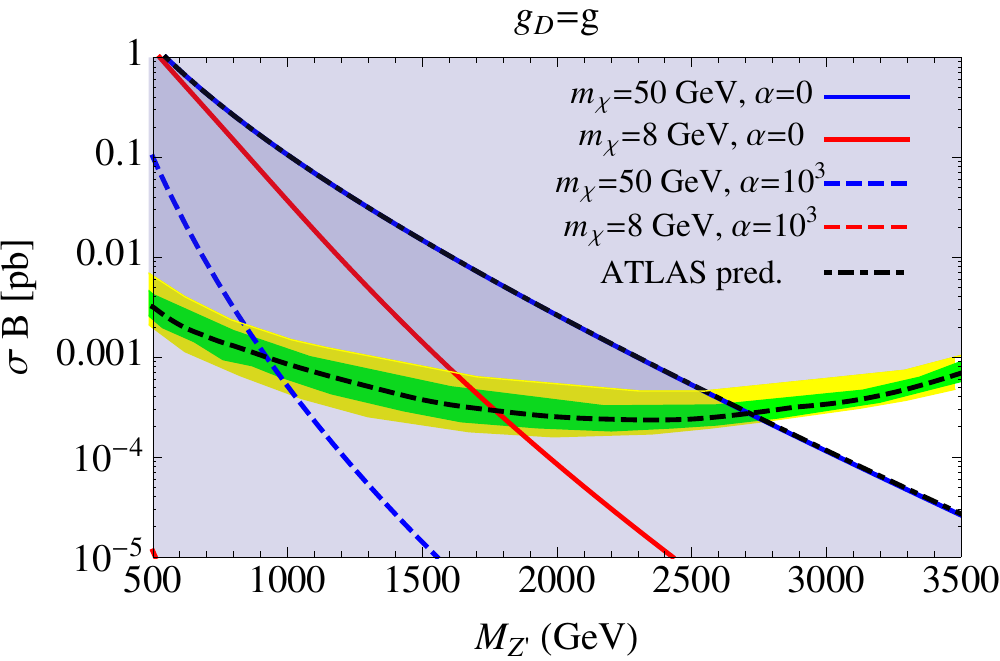}
              \caption{{\footnotesize
					Observational limit (black dashed lines) with $1-2 \sigma$ uncertainties (green-yellow band) on the production cross section times
					 branching ratio of decay into leptons for a $Z'$ prime particle as reported by ATLAS collaboration. The blue and red lines represent
					  the LUX exclusion limits for $g_D=g$ and two values of the DM mass, namely $m_\chi=8$ and 50 GeV, assuming pure vectorial (solid lines), namely 
					  $\alpha=0$, and substantially pure axial (dashed lines), $\alpha=10^3$.
					  The black dash-dotted line represent the prediction of $\sigma B$ for the SSM. 
					   As discussed in the text this coincides with the LUX exclusion for $m_\chi=50$ GeV and pure vectorial couplings since for this
					    value of the mass only a very small invisible branching ratio is allowed.  		
}}
\label{Fig:sigmavsMz1}
\end{center}
\end{figure}

\begin{figure}[t]
    \begin{center}
    \includegraphics[width=4.3in]{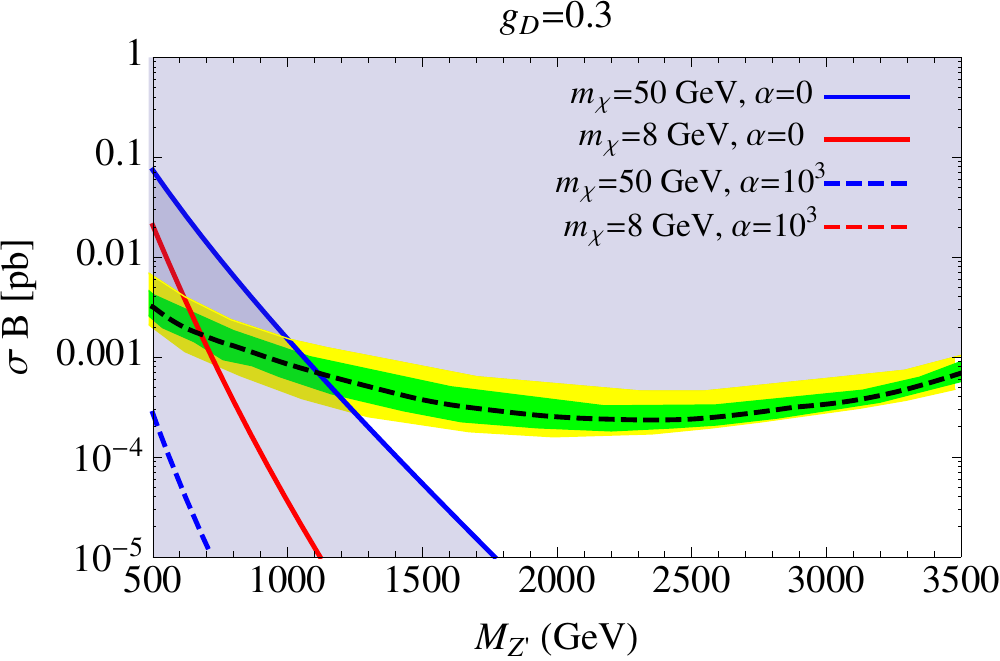}
              \caption{{\footnotesize
        The same as Fig.~(\ref{Fig:sigmavsMz1}) for $g_D=0.3$. }}
\label{Fig:sigmavsMz2}
\end{center}
\end{figure}

We compare the exclusion limits from the ATLAS collaboration with the ones set by LUX, expressed 
in terms of the resonant dilepton production and of the $Z'$ mass and for  $g_D = g$ and $g_D=0.3$ in figures (\ref{Fig:sigmavsMz1}) and (\ref{Fig:sigmavsMz2}) 
respectively.
 One clearly understands from  these figures the complementarity between the LHC and direct detection searches: for a given $Z'$ mass, the LHC sets an upper limit on the 
production cross section while, in our framework, LUX sets a lower limit. This is because a smaller dilepton production corresponds to larger invisible
 branching ratio, and thus larger coupling to DM. We then understand easily that the LUX constraint gets weaker if the coupling $g_D$ is smaller. 
A similar conclusion holds if there is substantial axial coupling.

The dependence of the direct detection exclusion limits on the dark mass is also illustrated in figures  (\ref{Fig:sigmavsMz1}) and (\ref{Fig:sigmavsMz2}).
For light dark matter ($8$ GeV for example), the constraints from direct detection are weaker, thus allowing more invisible branching, and a weaker limit on the $Z'$ mass. 
On the contrary, for a $50$ GeV dark matter, the LUX experiment substantially disfavors the presence of an invisible branching ratio, in the case of purely vectorial couplings, such that 
the LUX line practically overlaps with the prediction of the dilepton production cross section for the SSM without dark matter. In all cases, the allowed parameter range, 
for a given $Z'$ realization and DM mass, corresponds to the region which is simultaneously below the LHC observational limit (black dashed line in the plots)
 and above the corresponding LUX exclusion line. 

\begin{figure}[t]
    \begin{center}
    \includegraphics[width=4.3in]{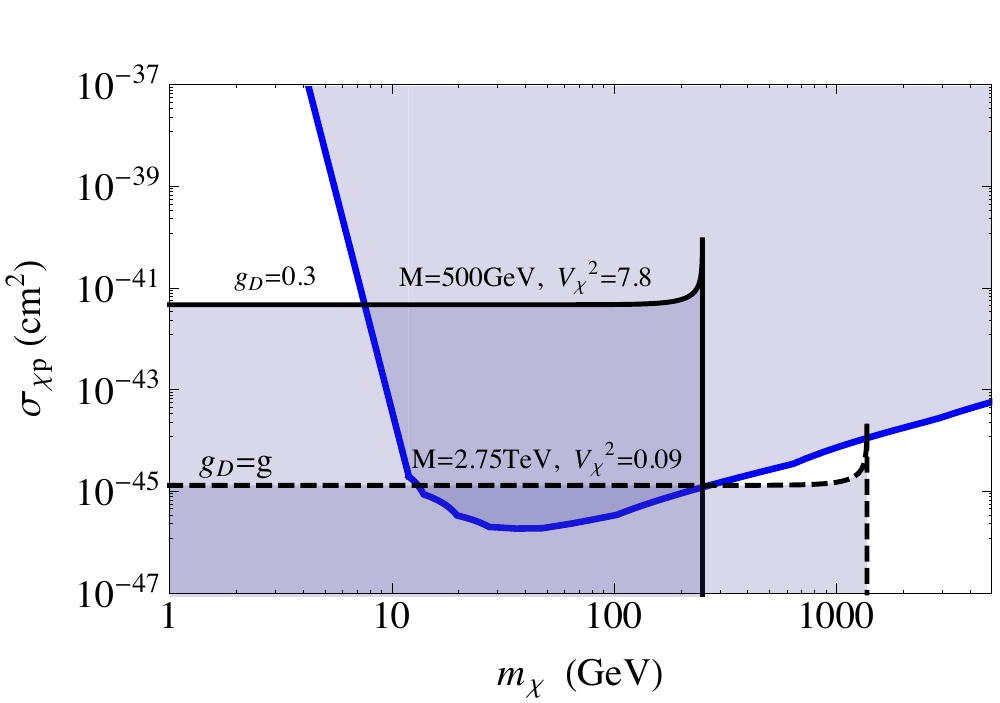}
              \caption{
{\footnotesize
       Constraints from dilepton searches shown in the plane 
$\sigma_{SI}-m_\chi$
 together with the LUX exclusion limits. See text for details.  }
}
\label{Fig:sigmaSImChi}
\end{center}
\end{figure}

It is of interest to present the above constraints  in the plane $\sigma_{SI}-m_\chi$, see figure~(\ref{Fig:sigmaSImChi}). 
There we show again the fact that, for a given $Z'$ mass, dilepton searches put a lower bound on the $Z'$ coupling to dark matter (which is quantified by 
$V_\chi$ for fixed $g_D$). By the same token, a larger $V_\chi$ leads to a larger SI cross section, which may be excluded by direct searches. This is illustrated for two different ($M_{Z'}$ ,$g_D$) pairs. The features in this figure are pretty clear. Obviously there is no constraint from {resonant production} for $M_{Z'} > 2 m_\chi$, since in this case the $Z'$ is fully visible. 
The spike for $m_{\chi} \lesssim M_{Z'}/2$ is due to phase-space suppression, which requires a larger value of the coupling to dark matter to be compatible with dilepton searches (the values of $V_\chi$ quoted in the figure are for $m_{\chi} \ll M_{Z'}/2$). Of course there are no constraints from dileptons if the $Z'$ is too heavy to be produced at the LHC (e.g. for $M_{Z'} \gsim 2.8$ TeV for $g_D = g$), but a lighter $Z'$ is {\it a priori} possible provided it has a larger coupling to an invisible sector. However this coupling cannot be arbitrarily large. Imposing $g_D^2 V_D^2 \lesssim 4 \pi$ gives a lower bound on the $Z'$ mass, which we will refer to as a unitarity bound in the sequel. 

\begin{figure}[t]
    \begin{center}
      \includegraphics[width=4.3in]{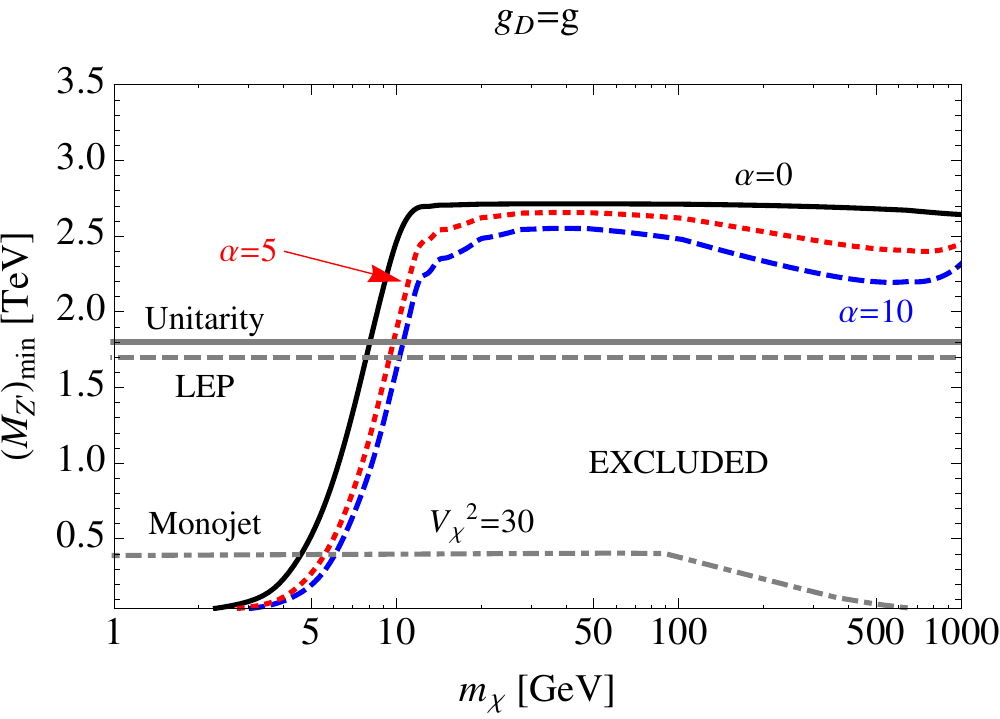}
              \caption{{\footnotesize
         Minimum values of $M_{Z'}$ for the SSM when combining LUX and ATLAS bounds in the case of the SSM for different nature of the dark matter coupling
         to the $Z'$ ($\alpha = 0, 5$ and 10). Also shown is the bound from LEP (dashed horizontal line), the perturbativity bound on the coupling to dark matter (solid horizontal line) and the limit from monojet plus missing energy at the LHC. The points below the lines are excluded.}}
\label{Fig:combined}
\end{center}
\end{figure}

Combining all our preceding results, we give in the figure (\ref{Fig:combined}) the parameter space allowed when combining ATLAS and LUX limits
in the plane ($m_\chi$, $M_{Z'}$) and in the case of the SSM. All the points below the lines are excluded by our analysis.
We see that a large region does not survive the constraints, as expected from the discussion above.  
For instance one may conclude from this figure that it is excluded that a $Z'$ lighter than about 2--3 TeV couples to a dark matter candidate with mass $\gtrsim 50$ GeV.
This rather strong conclusion holds for any natural $U(1)'$ model studied here, since we have considered variations of order unity for the vector and axial couplings $V^f$ and $A^f$, see Table I.  
The features in  Fig.(\ref{Fig:combined}) are otherwise easy to understand. The plateau for $\alpha=0$ corresponds to the current bound on $M_{Z'}$ set by the LHC experiments (here ATLAS), while the rise  near $M_{Z'}\simeq 2 m_\chi$ for the heaviest DM candidates is simply due to a threshold effect as $Br_\chi \rightarrow 0$.
For light dark matter ($m_{\chi}\lesssim$ 10 GeV) direct detection experiments allow for a larger invisible branching, and thus the LHC bound on the $Z'$ mass can be much weaker.  Similarly, for
 non-zero $\alpha$, the constraints on invisible branching from LUX are weaker, and so the mass of the $Z'$ may also be smaller. 
 
 In the region of the plots where the constraints from both direct detection and dileptons resonance are weak, one may expect that other limits become relevant.
In the plot, we show the LEP limits on (non-resonant) dileptons production (dashed horizontal line), which for $g_D = g$ in the SSM model, corresponds to $M_{Z'} \gtrsim 1.7$ TeV \cite{Langacker:2008yv,Alcaraz:2006mx} (for other couplings, one may use the fact that this limits scales likes $g_D^4/M_{Z'}^4$). We have also reported the unitarity bound, $M_{Z'} \gtrsim 1.8$ TeV (solid horizontal line), which for the SSM is more constraining than the LEP limit\footnote{{We remind that this unitarity limit comes from the dileptons analysis, see previous section. }}. 
 Notice that these two constraints are particularly strong because  we have chosen $g_D = g$ and could be relaxed by considering a weaker couplings of the $Z'$ to leptons.
We finally report our estimate of the limit from monojet searches, $M_{Z'} \gtrsim 0.4$ TeV (dot-dashed line) for $g_D^2 V_D^2 = 4 \pi$. This bound is weaker than the two other bounds. However,  because of the possibility of a resonant $Z'$, notice that our estimate is stronger than the ones reported in {\it e.g.} ~\cite{Aad:2013oja,Chatrchyan:2012me}, which rely on an effective operator analysis.\footnote{Analogous constraints come from mono-photons plus missing energy, which are however weaker than the monojet ones.} 
Moreover, the limit from monojet searches depends on the interplay between the coupling of the $Z'$ with DM and leptons, so that the monojet constraint could be relevant in other realizations and should not be {\it a priori} neglected.

\section*{Conclusion and Prospects}

We have considered a very generic scenario in which a $Z'$ gauge boson couples both to the Standard Model fermions and to a dark matter candidate, which belongs to some invisible sector. 
We have encompassed  many realizations existing in the literature by using a simple parametrization of the couplings of the Standard Model and dark sector 
to the $Z'$. We have studied the complementarity between the searches for $Z'$ at LHC and those for dark matter by direct detection experiments. For the 
sake of illustration, we have confronted on one hand the constraints on the resonant production of dileptons at the LHC  and on the other hand the exclusion limits 
on spin independent elastic scattering set by the LUX collaboration. 
Our main point is illustrated in Figs. (\ref{Fig:sigmavsMz1}-\ref{Fig:sigmaSImChi}), where we show how
 a $Z'$ with substantial invisible branching ratio is constrained by the current direct detection experiments, and LUX 
 in particular. 
The model dependent character of this statement is illustrated through the parametric dependence in an universal coupling parameter $g_D$, 
 and a parameter noted $\alpha$ that parametrizes the sensitivity to spin independent scattering collisions in direct detection experiments (see Eq.\ref{Eq:alpha}).

  The final result of our analysis is shown in Fig.(\ref{Fig:combined}) in the case of a natural sequential model, in which all the Standard Model fermions  have the same couplings 
 as  that to the $Z$ boson. It shows that an electroweak scale dark matter candidate is essentially excluded in such a model, except for heavy $Z'$ ($M_{Z'} \gtrsim 2$ TeV).
 We also remark
 that we have been agnostic regarding the mechanism at the origin of the abundance of dark matter. Standard thermal production through annihilations mediated by $Z'$ 
 leads in general to an overabundance of dark matter in the parameter space allowed by our limits. The correct value of the dark matter relic density might be
 achieved thanks to additional interactions in the hidden sector or due to  a non-standard cosmological evolution.
In a forthcoming work, we will extend the phenomenological analysis of partially invisible $Z'$ models, including the dark matter abundance and several completions of the collider searches.

\noindent {\bf Note added}. During the completion of this work the paper \cite{Alves:2013tqa}, discussing the possible complementarity between detection strategies for dark $Z'$ models, has appeared. The authors consider the specific case  of a leptophobic $Z'$. Our analysis is more general and can be applied in principle to any $U(1)'$ extension of the standard model completed by a dark coupling.

\noindent {\bf Acknowledgements. }  The authors would like  to thank E. Dudas, K. Tarachenko, Ulitsa Dvoryanskaya
 and G. Belanger for very useful discussions.
This  work was supported by  the Spanish MICINN's
Consolider-Ingenio 2010 Programme  under grant  Multi-Dark {CSD2009-00064}.
G.A. and Y.M.  acknowledge partial support from the European Union FP7 ITN INVISIBLES (Marie
Curie Actions, PITN- GA-2011- 289442). Y.M. also acknowledges partial support from the ERC advanced grant Higgs@LHC. The work of M.T. and B.Z. is supported by the IISN, an ULB-ARC grant and by the Belgian Federal Science Policy through the Interuniversity Attraction Pole P7/37.


\vspace{1cm}


\begin{thebibliography}{99}











%%%%%%%%%%%%%%%%%%%%%%%%%%%%%%%%%%%%%%%%%%%%%%%%







%====================  INTRODUCTION  ===================================



\bibitem{Langacker:2008yv}
  P.~Langacker,
  %``The Physics of Heavy $Z^\prime$ Gauge Bosons,''
  Rev.\ Mod.\ Phys.\  {\bf 81} (2008) 1199
  [arXiv:0801.1345 [hep-ph]].

%\cite{Han:2013mra}
\bibitem{Han:2013mra} 
  T.~Han, P.~Langacker, Z.~Liu and L.~-T.~Wang,
  %``Diagnosis of a New Neutral Gauge Boson at the LHC and ILC for Snowmass 2013,''
  arXiv:1308.2738 [hep-ph].
  %%CITATION = ARXIV:1308.2738;%%
  %4 citations counted in INSPIRE as of 26 Dec 2013


%====================  STRING MIXING ===================================


%\cite{Cicoli:2011yh}
\bibitem{string}
  M.~Cicoli, M.~Goodsell, J.~Jaeckel and A.~Ringwald,
  %``Testing String Vacua in the Lab: From a Hidden CMB to Dark Forces in Flux
 % Compactifications,''
  arXiv:1103.3705 [hep-th];
  %%CITATION = ARXIV:1103.3705;%%
 M.~Goodsell, J.~Jaeckel, J.~Redondo and A.~Ringwald,
  %``Naturally Light Hidden Photons in LARGE Volume String Compactifications,''
  JHEP {\bf 0911} (2009) 027
  [arXiv:0909.0515 [hep-ph]];
 S.~A.~Abel, M.~D.~Goodsell, J.~Jaeckel, V.~V.~Khoze and A.~Ringwald,
  %``Kinetic Mixing of the Photon with Hidden U(1)s in String Phenomenology,''
  JHEP {\bf 0807} (2008) 124
  [arXiv:0803.1449 [hep-ph]];
  %%CITATION = JHEPA,0807,124;%%
  S.~Cassel, D.~M.~Ghilencea and G.~G.~Ross,
  %``Electroweak and Dark Matter Constraints on a Z' in Models with a Hidden
 % Valley,''
  Nucl.\ Phys.\  B {\bf 827} (2010) 256
  [arXiv:0903.1118 [hep-ph]];
  %%CITATION = NUPHA,B827,256;%%
M.~E.~Krauss, W.~Porod and F.~Staub,
  %``SO(10) inspired gauge-mediated supersymmetry breaking,''
  Phys.\ Rev.\ D {\bf 88} (2013) 015014
  [arXiv:1304.0769 [hep-ph]];
  %%CITATION = ARXIV:1304.0769;%%
%\bibitem{Hirsch:2012kv}
  M.~Hirsch, W.~Porod, L.~Reichert and F.~Staub,
  %``Phenomenology of the minimal supersymmetric $U(1)_{B-L}\times U(1)_R$ extension of the standard model,''
  Phys.\ Rev.\ D {\bf 86} (2012) 093018
  [arXiv:1206.3516 [hep-ph]];
S.~Andreas, M.~D.~Goodsell and A.~Ringwald,
  %``Dark Matter and Dark Forces from a Supersymmetric Hidden Sector,''
  Phys.\ Rev.\ D {\bf 87} (2013) 025007
  [arXiv:1109.2869 [hep-ph]].
  %%CITATION = ARXIV:1109.2869;%%


%--------------------------------------  Kinetic  -------------------------------------------------

\bibitem{Holdom}
B.~Holdom,
  %``Two U(1)'S And Epsilon Charge Shifts,''
  Phys.\ Lett.\  B {\bf 166}, 196 (1986).
  %%CITATION = PHLTA,B166,196;%%
  R.~Foot, X.~-G.~He,
  %``Comment on Z Z-prime mixing in extended gauge theories,''
  Phys.\ Lett.\  {\bf B267 } (1991)  509-512;
    R.~Foot, H.~Lew, R.~R.~Volkas,
  %``A Model with fundamental improper space-time symmetries,''
  Phys.\ Lett.\  {\bf B272 } (1991)  67-70;
  \bibitem{Feldman:2007wj}
  D.~Feldman, Z.~Liu and P.~Nath,
  %``The Stueckelberg Z' extension with kinetic mixing and milli-charged dark
 % matter from the hidden sector,''
  Phys.\ Rev.\  D {\bf 75} (2007) 115001
  [arXiv:hep-ph/0702123].
  %%CITATION = PHRVA,D75,115001;%%



\bibitem{Rizzo:1998ut}
  T.~G.~Rizzo,
  %``Gauge kinetic mixing and leptophobic $Z^\prime$ in E(6) and SO(10),''
  Phys.\ Rev.\  D {\bf 59} (1999) 015020
  [arXiv:hep-ph/9806397];
  %%CITATION = PHRVA,D59,015020;%%
    F.~del Aguila, M.~Masip and M.~Perez-Victoria,
  %``Physical parameters and renormalization of U(1)-a x U(1)-b models,''
  Nucl.\ Phys.\  B {\bf 456} (1995) 531
  [arXiv:hep-ph/9507455];
  %%CITATION = NUPHA,B456,531;%%
  B.~A.~Dobrescu,
  %``Massless gauge bosons other than the photon,''
  Phys.\ Rev.\ Lett.\  {\bf 94} (2005) 151802
  [arXiv:hep-ph/0411004];
  %%CITATION = PRLTA,94,151802;%%
  K.~R.~Dienes, C.~F.~Kolda and J.~March-Russell,
  %``Kinetic mixing and the supersymmetric gauge hierarchy,''
  Nucl.\ Phys.\  B {\bf 492} (1997) 104
  [arXiv:hep-ph/9610479];
  %%CITATION = NUPHA,B492,104;%%
  T.~Cohen, D.~J.~Phalen, A.~Pierce and K.~M.~Zurek,
  %``Asymmetric Dark Matter from a GeV Hidden Sector,''
  arXiv:1005.1655 [hep-ph].
  %%CITATION = ARXIV:1005.1655;%%
  
  %\cite{Chiang:2013kqa}
\bibitem{Chiang:2013kqa} 
  C.~-W.~Chiang, T.~Nomura and J.~Tandean,
  %``Nonabelian Dark Matter with Resonant Annihilation,''
  arXiv:1306.0882 [hep-ph].
  %%CITATION = ARXIV:1306.0882;%%
  %1 citations counted in INSPIRE as of 15 Jan 2014
  
  %\cite{Chiang:2012qa}
\bibitem{Chiang:2012qa} 
  C.~-W.~Chiang, T.~Nomura and J.~Tandean,
  %``Dark Matter and Higgs Boson in a Model with Discrete Gauge Symmetry,''
  Phys.\ Rev.\ D {\bf 87}, no. 7, 073004 (2013)
  [arXiv:1205.6416 [hep-ph]].
  %%CITATION = ARXIV:1205.6416;%%
  %2 citations counted in INSPIRE as of 15 Jan 2014






%---------------------------------------------  Direct kinetic  -----------------------------------------------


\bibitem{kindirect}
 M.~T.~Frandsen, F.~Kahlhoefer, S.~Sarkar and K.~Schmidt-Hoberg,
  %``Direct detection of dark matter in models with a light Z',''
  JHEP {\bf 1109} (2011) 128
  [arXiv:1107.2118 [hep-ph]];
  %%CITATION = ARXIV:1107.2118;%%
 Y.~Mambrini,
  %``The Kinetic dark-mixing in the light of CoGENT and XENON100,''
  JCAP {\bf 1009} (2010) 022
  [arXiv:1006.3318 [hep-ph]];
   E.~J.~Chun, J.~-C.~Park and S.~Scopel,
  %``Dark matter and a new gauge boson through kinetic mixing,''
  JHEP {\bf 1102} (2011) 100
  [arXiv:1011.3300 [hep-ph]];
  %%CITATION = ARXIV:1006.3318;%%
  H.~Davoudiasl and I.~M.~Lewis,
  %``Dark Matter from Hidden Forces,''
  arXiv:1309.6640 [hep-ph].


%--------------------------------------------  Indirect kinetic  ----------------------------------------------

\bibitem{kinindirect}
 Y.~Mambrini,
  %``The ZZ' kinetic mixing in the light of the recent direct and indirect dark matter searches,''
  JCAP {\bf 1107} (2011) 009
  [arXiv:1104.4799 [hep-ph]].
  %%CITATION = ARXIV:1104.4799;%%


%------------------------------------------  kinetic + scalar  -------------------------------------------------

\bibitem{kinscalar}
  R.~Ramos and M.~Sher,
  %``The Dark Z and Charged Higgs Decay,''
  arXiv:1312.0013 [hep-ph];
  %%CITATION = ARXIV:1312.0013;%%
 G.~Belanger, A.~Goudelis, J.~-C.~Park and A.~Pukhov,
  %``Isospin-violating dark matter from a double portal,''
  arXiv:1311.0022 [hep-ph].
  %%CITATION = ARXIV:1311.0022;%%
  %1 citations counted in INSPIRE as of 28 Dec 2013




%----------------------------------------  EXPERIMENTS   -----------------------------------------------------------




%\cite{Aad:2012hf}
\bibitem{Aad:2012hf}
  G.~Aad {\it et al.}  [ATLAS Collaboration],
  %``Search for high-mass resonances decaying dilepton final states in pp collisions at s**(1/2) = 7-TeV with the ATLAS detector,''
  JHEP {\bf 1211} (2012) 138
  [arXiv:1209.2535 [hep-ex]].
  %%CITATION = ARXIV:1209.2535;%%
  %39 citations counted in INSPIRE as of 25 Dec 2013


%\cite{Chatrchyan:2012oaa}
\bibitem{Chatrchyan:2012oaa}
  S.~Chatrchyan {\it et al.}  [CMS Collaboration],
  %``Search for heavy narrow dilepton resonances in $pp$ collisions at $\sqrt{s}=7$ TeV and $\sqrt{s}=8$ TeV,''
  Phys.\ Lett.\ B {\bf 720} (2013) 63
  [arXiv:1212.6175 [hep-ex]].
  %%CITATION = ARXIV:1212.6175;%%
  %27 citations counted in INSPIRE as of 22 Dec 2013

\bibitem{Aprile:2012nq}
  E.~Aprile {\it et al.}  [XENON100 Collaboration],
  %``Dark Matter Results from 225 Live Days of XENON100 Data,''
  Phys.\ Rev.\ Lett.\  {\bf 109} (2012) 181301
  [arXiv:1207.5988 [astro-ph.CO]].
  %%CITATION = ARXIV:1207.5988;%%
  %466 citations counted in INSPIRE as of 15 Dec 2013

\bibitem{Akerib:2013tjd}
  D.~S.~Akerib {\it et al.}  [LUX Collaboration],
  %``First results from the LUX dark matter experiment at the Sanford Underground Research Facility,''
  arXiv:1310.8214 [astro-ph.CO].
  %%CITATION = ARXIV:1310.8214;%%
  %43 citations counted in INSPIRE as of 15 Dec 2013


\bibitem{Archambault:2012pm}
  S.~Archambault {\it et al.}  [PICASSO Collaboration],
  %``Constraints on Low-Mass WIMP Interactions on $^{19}F$ from PICASSO,''
  Phys.\ Lett.\ B {\bf 711} (2012) 153
  [arXiv:1202.1240 [hep-ex]].
  %%CITATION = ARXIV:1202.1240;%%
  %65 citations counted in INSPIRE as of 15 Dec 2013

\bibitem{Behnke:2012ys}
  E.~Behnke {\it et al.}  [COUPP Collaboration],
  %``First Dark Matter Search Results from a 4-kg CF$_3$I Bubble Chamber Operated in a Deep Underground Site,''
  Phys.\ Rev.\ D {\bf 86} (2012) 052001
  [arXiv:1204.3094 [astro-ph.CO]].
  %%CITATION = ARXIV:1204.3094;%%
  %62 citations counted in INSPIRE as of 15 Dec 2013








%--------------------------------------  LHC Z'  ----------------------------------------

\bibitem{zplhc}
 M.~T.~Frandsen, F.~Kahlhoefer, A.~Preston, S.~Sarkar and K.~Schmidt-Hoberg,
  %``LHC and Tevatron Bounds on the Dark Matter Direct Detection Cross-Section for Vector Mediators,''
  JHEP {\bf 1207} (2012) 123
  [arXiv:1204.3839 [hep-ph]];
  V.~Barger, D.~Marfatia and A.~Peterson,
  %``LHC and Dark Matter Signals of Z' Bosons,''
  Phys.\ Rev.\ D {\bf 87} (2013) 015026
  [arXiv:1206.6649 [hep-ph]];
  %%CITATION = ARXIV:1206.6649;%%
  %2 citations counted in INSPIRE as of 28 Dec 2013
 E.~Dudas, L.~Heurtier, Y.~Mambrini and B.~Zaldivar,
  %``Extra U(1), effective operators, anomalies and dark matter,''
  JHEP {\bf 1311} (2013) 083
  [arXiv:1307.0005 [hep-ph]];
  %%CITATION = ARXIV:1307.0005;%%
D.~Feldman, P.~Fileviez Perez and P.~Nath,
  %``R-parity Conservation via the Stueckelberg Mechanism: LHC and Dark Matter Signals,''
  JHEP {\bf 1201} (2012) 038
  [arXiv:1109.2901 [hep-ph]];
 %%CITATION = ARXIV:1107.0771;%%
H.~-S.~Lee and Y.~Li,
  %``Identifying Sneutrino Dark Matter: Interplay between the LHC and Direct Search,''
  Phys.\ Rev.\ D {\bf 84}, 095003 (2011)
  [arXiv:1107.0771 [hep-ph]].
  %%CITATION = ARXIV:1107.0771;%%
  %2 citations counted in INSPIRE as of 12 Jan 2014


%-----------------------------------------  Direct Z' ------------------------------------
\bibitem{zpdirect}
 P.~Gondolo, P.~Ko and Y.~Omura,
  %``Light dark matter in leptophobic Z' models,''
  Phys.\ Rev.\ D {\bf 85} (2012) 035022
  [arXiv:1106.0885 [hep-ph]];
  N.~Fornengo, P.~Panci and M.~Regis,
  %``Long-Range Forces in Direct Dark Matter Searches,''
  Phys.\ Rev.\ D {\bf 84} (2011) 115002
  [arXiv:1108.4661 [hep-ph]].




%=========================== SPIN  =========================

\bibitem{Airapetian:2007mh}
  A.~Airapetian {\it et al.}  [HERMES Collaboration],
  %``Precise determination of the spin structure function g(1) of the proton, deuteron and neutron,''
  Phys.\ Rev.\ D {\bf 75} (2007) 012007
  [hep-ex/0609039].
  %%CITATION = HEP-EX/0609039;%%
  %256 citations counted in INSPIRE as of 28 Dec 2013



%==========================  DD   ============================

\bibitem{dd}
 G.~Belanger, F.~Boudjema, A.~Pukhov and A.~Semenov,
  %``Dark matter direct detection rate in a generic model with micrOMEGAs 2.2,''
  Comput.\ Phys.\ Commun.\  {\bf 180} (2009) 747
  [arXiv:0803.2360 [hep-ph]];
  %%CITATION = ARXIV:0803.2360;%%
  %361 citations counted in INSPIRE as of 28 Dec 2013
 G.~Jungman, M.~Kamionkowski and K.~Griest,
  %``Supersymmetric dark matter,''
  Phys.\ Rept.\  {\bf 267} (1996) 195
  [hep-ph/9506380];
  %%CITATION = HEP-PH/9506380;%%
 D.~G.~Cerdeno and A.~M.~Green,
  %``Direct detection of WIMPs,''
  In *Bertone, G. (ed.): Particle dark matter* 347-369
  [arXiv:1002.1912 [astro-ph.CO]].
  %%CITATION = ARXIV:1002.1912;%%
  
  

%==========================  LHC exp bis  =============================


%==================== LHC exp ==========================

%\cite{ATLAS:2012pu}
\bibitem{ATLAS:2012pu}
  G.~Aad {\it et al.}  [ATLAS Collaboration],
  %``ATLAS search for new phenomena in dijet mass and angular distributions using $pp$ collisions at $\sqrt{s}=7$ TeV,''
  JHEP {\bf 1301} (2013) 029
  [arXiv:1210.1718 [hep-ex]].
  %%CITATION = ARXIV:1210.1718;%%
  %48 citations counted in INSPIRE as of 25 Dec 2013


%\cite{Chatrchyan:2013qha}
\bibitem{Chatrchyan:2013qha}
  S.~Chatrchyan {\it et al.}  [CMS Collaboration],
  %``Search for narrow resonances using the dijet mass spectrum in pp collisions at sqrt(s) = 8 TeV,''
  Phys.\ Rev.\ D {\bf 87} (2013) 114015
  [arXiv:1302.4794 [hep-ex]].
  %%CITATION = ARXIV:1302.4794;%%
  %31 citations counted in INSPIRE as of 25 Dec 2013

%\cite{Aad:2013oja}
\bibitem{Aad:2013oja}
  G.~Aad {\it et al.}  [ATLAS Collaboration],
  %``Search for dark matter in events with a hadronically decaying W or Z boson and missing transverse momentum in pp collisions at $\sqrt{s}$=8 TeV with the ATLAS detector,''
  arXiv:1309.4017 [hep-ex].
  %%CITATION = ARXIV:1309.4017;%%
  %8 citations counted in INSPIRE as of 25 Dec 2013


%\cite{Chatrchyan:2012me}
\bibitem{Chatrchyan:2012me}
  S.~Chatrchyan {\it et al.}  [CMS Collaboration],
  %``Search for dark matter and large extra dimensions in monojet events in $pp$ collisions at $\sqrt{s}=7$ TeV,''
  JHEP {\bf 1209} (2012) 094
  [arXiv:1206.5663 [hep-ex]].
  %%CITATION = ARXIV:1206.5663;%%
  %115 citations counted in INSPIRE as of 25 Dec 2013
  
%\cite{Fox:2012ru}
\bibitem{Fox:2012ru}
  P.~J.~Fox and C.~Williams,
  %``Next-to-Leading Order Predictions for Dark Matter Production at Hadron Colliders,''
  Phys.\ Rev.\ D {\bf 87} (2013) 054030
  [arXiv:1211.6390 [hep-ph]].
  %%CITATION = ARXIV:1211.6390;%%
  %18 citations counted in INSPIRE as of 14 Feb 2014


  %\cite{Feng:2012jn}
\bibitem{Feng:2012jn} 
  W.~-Z.~Feng, P.~Nath and G.~Peim,
  %``Cosmic Coincidence and Asymmetric Dark Matter in a Stueckelberg Extension,''
  Phys.\ Rev.\ D {\bf 85}, 115016 (2012)
  [arXiv:1204.5752 [hep-ph]].
  %%CITATION = ARXIV:1204.5752;%%
  %17 citations counted in INSPIRE as of 07 Jan 2014


 



%\cite{Alcaraz:2006mx}
\bibitem{Alcaraz:2006mx}
  J.~Alcaraz {\it et al.}  [ALEPH and DELPHI and L3 and OPAL and LEP Electroweak Working Group Collaborations],
  %``A Combination of preliminary electroweak measurements and constraints on the standard model,''
  hep-ex/0612034.
  %%CITATION = HEP-EX/0612034;%%
  %320 citations counted in INSPIRE as of 14 Feb 2014












%====================== INTRODUCTION  ================================



	%\cite{Accomando:2010fz}
\bibitem{Accomando:2010fz} 
  E.~Accomando, A.~Belyaev, L.~Fedeli, S.~F.~King and C.~Shepherd-Themistocleous,
  %``Z' physics with early LHC data,''
  Phys.\ Rev.\ D {\bf 83}, 075012 (2011)
  [arXiv:1010.6058 [hep-ph]].
  %%CITATION = ARXIV:1010.6058;%%
  %51 citations counted in INSPIRE as of 26 Dec 2013


%==============Brazilian paper============================
%\cite{Alves:2013tqa}
\bibitem{Alves:2013tqa} 
  A.~Alves, S.~Profumo and F.~S.~Queiroz,
  %``The Dark Z' Portal: Direct, Indirect and Collider Searches,''
  arXiv:1312.5281 [hep-ph].
  %%CITATION = ARXIV:1312.5281;%%
	

%%%%%%%%%%%%%%%%%%%%%%%%%%%%%%%new references%%%%%%%%%%%%%%%%%%%%%%
  

  





\end{thebibliography}
\end{document}